\documentclass[prd, preprint, nofootinbib, showpacs]{revtex4}
\usepackage{epsfig,graphicx,amsmath,amssymb}
\usepackage{amsfonts}
\usepackage{latexsym}
\usepackage{color}

\newcommand{\dis}[1]{\begin{equation}\begin{split}#1\end{split}\end{equation}}
\newcommand{\be}{\begin{equation}}
	\newcommand{\ee}{\end{equation}}
\def\bea{\begin{eqnarray}}
	\def\eea{\end{eqnarray}}

\newcommand{\eq}[1]{Eq.~(\ref{#1})}

\newcommand{\bfrac}[2]{{\left(\frac{#1}{#2} \right)  }}\newcommand{\VEV}[1]{\langle #1 \rangle}

\newcommand{\Mp}{M_P}

\newcommand\tev{\,{\rm TeV}}
\newcommand\gev{\,{\rm GeV}}
\newcommand\mev{\,{\rm MeV}}
\newcommand\kev{\,{\rm keV}}

\newcommand\cm{\,{\rm cm}}

\newcommand{\nus}{\nu_{s}} 
\newcommand{\nusj}{{\nu_{sj}}} 
\newcommand{\nush}{{\nu_{sh}}} 

\newcommand{\ms}{m_{\nu_s}} 
\newcommand{\msj}{m_{\nu_{sj}}} 
\newcommand{\msi}{m_{\nu_{si}}} 

\begin{document}

\title{Sterile neutrino dark matter with dipole interaction}

\author{Wonsub Cho}
\email{sub526@skku.edu}

\author{Ki-Young Choi}
\email{kiyoungchoi@skku.edu}
\affiliation{Department of Physics, Sungkyunkwan University,  2066, Seobu-ro, Jangan-gu, Suwon-si, Gyeong Gi-do, 16419 Korea}

\author{Osamu Seto}
\email{seto@particle.sci.hokudai.ac.jp}
\affiliation{Institute for the Advancement of Higher Education, Hokkaido University, Sapporo 060-0817, Japan}
\affiliation{Department of Physics, Hokkaido University, Sapporo 060-0810, Japan}

%

\begin{abstract}
We consider the possibility of the lightest sterile neutrino dark matter which has dipole interaction with heavier sterile neutrinos. The lifetime can be long enough to be a dark matter candidate without violating other constraints and the correct amount of relic abundance can be produced in the early Universe. We find that a sterile neutrino with the mass of around MeV  and the dimension-five non-renormalisable dipole interaction suppressed by $\Lambda_5 \gtrsim 10^{15}$ GeV can be a good candidate of dark matter, while heavier sterile neutrinos with masses of the order of GeV can explain the active neutrino oscillations.
%
\end{abstract}

\pacs{}

\preprint{EPHOU-21-012}

\vspace*{3cm}
\maketitle


\section{Introduction}
\label{sec:intro}

In the context of the standard model (SM) of particle physics, the oscillations between neutrino flavor species and the invisible component of the mass in the Universe cannot find any solution in itself. Both problems require new physics with new particles or new interactions beyond the standard model (BSM).

The simplest explanation of neutrino oscillation phenomena is that neutrinos have tiny masses, which can be explained by the see-saw mechanism with the introduction of heavy right-handed Majorana neutrinos~\cite{Minkowski:1977sc,Yanagida:1979as,GellMann:1980vs,Mohapatra:1979ia}. 
To explain two squared mass differences measured from the solar and atmospheric neutrino oscillations, only two kinds of right-handed neutrinos are enough. If three generations of right-handed neutrinos are assumed like other SM fermions, 
the remaining one right-handed neutrino can be practically decoupled from neutrino oscillation and free from the observations, thus it can be a candidate for sterile neutrino dark matter~\cite{Dodelson:1993je,Dolgov:2000ew,Asaka:2005an}.

Sterile neutrino dark matter with keV-scale mass has been studied thoroughly, since those can be produced in the early Universe through the simplest mechanism of Dodelson-Widrow (DW)~\cite{Dodelson:1993je} with proper amount for dark matter, and those can be warm dark matter with a significant free streaming scale that could help to relax the problems of the cold dark matter. However, the sterile neutrino DM generated by DW mechanism are now severely constrained from the observations in X-ray~\cite{Boyarsky:2005us,Boyarsky:2006fg,Boyarsky:2006ag,Boyarsky:2007ay,Yuksel:2007xh} and structure formation~\cite{Seljak:2006qw,Boyarsky:2008xj,Perez:2016tcq}, even if the hadronic uncertainties are taken into account~\cite{Asaka:2006nq}. For this, see a recent review Refs.~\cite{Adhikari:2016bei,Boyarsky:2018tvu}. That has brought other mechanisms for the production of dark matter such as the resonant oscillation production~\cite{Shi:1998km} or other non-thermal production mechanisms. 
Nonthermal production mechanisms of sterile neutrino DM include the decay of an extra singlet scalar~\cite{Shaposhnikov:2006xi,Kusenko:2006rh}, scatterings through new mediators in the thermal bath without reaching thermal equilibrium~\cite{Khalil:2008kp,Kaneta:2016vkq,Biswas:2016bfo,Seto:2020udg,DeRomeri:2020wng,Lucente:2021har,Belanger:2021slj} which is recently called ``freeze-in''~\cite{Hall:2009bx}. For a review on freeze-in scenarios, see, e.g., Refs.~\cite{Baer:2014eja,Shakya:2015xnx}. 

The possibility of the neutrino magnetic moment has been studied from long time ago~\cite{Cowan:1957pp,Bernstein:1963jp,Kim:1974xx}. While the magnetic moments between the active neutrinos can arise from the dimension-six operators, those between the gauge singlet right-handed neutrinos can arise from the dimension-five operator~\cite{Aparici:2009fh}. Considering that both operators are induced by the same new physics, the dimension-five operator is much less suppressed and may leave observable effects of high scale physics than the dimension-six operator. Provided that neutrino masses are generated through seesaw mechanism, neutrinos are Majorana particle, being self-conjugate, hence the magnetic moments act as the transition moments indeed. This transition induces the decay of heavier neutrinos to a lighter neutrino emitting a photon and various new scattering channels. Even if the effects are highly suppressed and negligible in terrestrial experiments, those may contribute to the decay and production of sterile neutrino dark matter. 

In this paper, we explore the possibility of the lightest sterile neutrino as dark matter which interact with other sterile neutrinos and the SM particles through the dipole interaction.
The sterile neutrino DM can decay into a photon and an active neutrino through the dipole operator, even if its Yukawa coupling is vanishing.
The bound on the sterile neutrino DM lifetime constrains the scale of the dipole interaction as well as the mixing between left-handed and other sterile neutrinos.
On the other hand, through this dipole interaction, sterile neutrinos are produced by the scattering processes in thermal plasma in the early Universe. We find that a sterile neutrino with the mass of around MeV and the dimension-five non-renormalisable dipole interaction suppressed by $\Lambda_5 \gtrsim 10^{15}$ GeV can be a good candidate for dark matter, while heavier sterile neutrinos with masses of the order of GeV can explain the active neutrino oscillations.

This paper is organized as follows. After we describe Langrangian of the model and give mass eigenstates and those interaction vertices in Sec.~\ref{sec:model}, we examine the scenarios of sterile neutrino DM
 by estimating its life time and DM abundance generated in the early Universe in Sec.~\ref{sec:dm}. 
 We mention the observation signatures in Sec.~\ref{sec:test}, and then conclude in Sec.~\ref{sec:concl}.

\section{Model}
\label{sec:model}
In this section, after we introduce the model, we derive the Lagrangian in the mass eigenstates and summarize their basic properties. The interactions of the sterile neutrinos will be used in the following sections.

We consider a model with Lagrangian including standard model (SM) part $ \mathcal{L}_{\mathrm{SM}} $ and additional one containing three right-handed neutrinos $\mathcal{L}_{\nu_R}$ as
\begin{align}
	\mathcal{L} = & \mathcal{L}_{\mathrm{SM}} + \mathcal{L}_{\nu_R},
\end{align}
where
\begin{align}
	\mathcal{L}_{\nu_R} = &
	- \frac{1}{2}\overline{\nu_{Ri}^c} M_{{\nu_R}_{ij}} \nu_{Rj}
	+ y_{\nu\alpha i} \overline{L_{\alpha}}\widetilde{\Phi}\nu_{R i}
	+ C_{ij} \overline{\nu_{Ri}^c}[\gamma^{\mu},\gamma^{\nu}] \nu_{Rj} B_{\mu\nu} + \mathrm{h.c.}.
\end{align}
Here, $B_{\mu\nu}$ is the gauge field strength of $U(1)_Y$ gauge field $B_{\mu}$,
$L_{\alpha}$ are lepton doublets with $\alpha$ flavor, $\Phi$ is the Higgs doublet 
and forms $\widetilde{\Phi} = \epsilon \Phi^*$ with
the superscript $c$ and $*$ for charge conjugation.
The Majorana mass of $\nu_R$ are taken to be diagonal, real and positive as
$M_{{\nu_R}_{ij}} = \mathrm{diag}(M_{{\nu_R}_1},M_{{\nu_R}_2},M_{{\nu_R}_3})$
without loss of generality.
We also note that the dipole interaction above is the most general form for $\nu_R$
because of the identity $\gamma_5 P_R = P_R$. 
The dipole interaction is a dimension-$5$ operator and the coupling $C_{ij}=\frac{c_{ij}}{\Lambda_5}$ is suppressed by a high energy scale $\Lambda_5$ with an anti-symmetric coupling $c_{ij}$ of the order of unity.

After the electroweak symmetry breaking, with the vacuum expectation value (VEV) $v=246$ GeV
of the SM Higgs field $\Phi$, the $B$ gauge boson and neutrinos are decomposed into the mass basis as
\begin{align}
	B_{\mu} =       & c_W A_{\mu} - s_W Z_{\mu}, \label{Eq:B}                          \\
	\nu_{L\alpha} = & U_{\alpha i}\nu_i + \Theta_{\alpha i} \nu_{si}^c, \label{Eq:nuL} \\
	\nu_{Ri}^c =    & - (\Theta^{\dagger} U)_{ij}\nu_j + \nu_{si}^c, \label{Eq:nuR}
\end{align}
where $c_W$ $(s_W)$ is cosine (sine) of the Weinberg angle, and $A_\mu$ and $Z_\mu$ are the photon and $Z$-boson.
$\nu_i$ and $\nu_{si}$ are the mass eigenstates of light active neutrinos with the mass eigenvalues $m_{\nu_i}$ and those of sterile neutrinos with the mass eigenvalues $m_{\nu_{si}}$, respectively.
Here, $i$ or $j$ runs from $1$ to $3$, and we have $m_{\nu_{sj}} \simeq M_{\nu_{Rj}}$ due to the mass hierarchy.
$m_{\nu_i}$ are eigenvalues of the mass matrix of the light neutrinos, which is generated through the seesaw mechanism~\cite{Minkowski:1977sc,Yanagida:1979as,GellMann:1980vs,Mohapatra:1979ia} as 
\begin{equation}
	m_{\nu} \simeq -m_D \frac{1}{M_{\nu_R}} m_D^T= - \Theta M_{\nu_R} \Theta^T ,
\end{equation}
in the flavor basis. Those are related as $m_{\nu}^\mathrm{diag}= \mathrm{diag}(m_1,m_2,m_3) =U^{\dagger} m_{\nu} U^* $ through the PMNS matrix $U$~\cite{Pontecorvo:1957qd,Maki:1962mu}.
The mixing between left- and right-handed neutrinos are parameterized by the mixing matrix
\begin{equation}
\Theta = m_DM^{-1}_{\nu_R} \ll 1,
	\label{Theta}
\end{equation}
with the Dirac mass $(m_D)_{\alpha i} = y_{\nu\alpha i}v/\sqrt{2}$.
Naively, the typical magnitude of the left-right mixing can be estimated as
\begin{equation}
	\Theta^2 \sim \frac{m_{\nu}}{M_{\nu_R} } .
	\label{Theta2}
\end{equation}

For the lightest sterile neutrino being the DM candidate, its lifetime should be long enough compared to the age of the Universe. This is obtained when
$y_{\nu\alpha 1}$ is practically zero so that we consider the Yukawa coupling
\begin{equation}
	y_{\nu} = \left(
	\begin{array}{lll}
			0 & y_{\nu e2}    & y_{\nu e3}    \\
			0 & y_{\nu\mu 2}  & y_{\nu\mu 3}  \\
			0 & y_{\nu\tau 2} & y_{\nu\tau 3} \\
		\end{array}
	\right) .
	\label{Yukawamatrix}
\end{equation}
Even with this vanishing Yukawa coupling between the lightest sterile neutrino and the active neutrinos, it is still possible to reproduce the light neutrino masses
to explain the observed neutrino oscillations~\cite{Asaka:2005an}.
Following the parametrization of Casas and Ibarra ~\cite{Casas:2001sr},
the Dirac mass term or the neutrino Yukawa coupling can be expressed as
\begin{equation}
	y_{\nu\alpha i} \frac{v}{\sqrt{2}}  = i U (m^\mathrm{diag}_{\nu}){}^{1/2} R (M_{\nu_R})^{1/2},
\label{eq:NuYukawa}
\end{equation}
with $R$ being a complex orthogonal matrix with $RR^T=1$.
For the normal ordering of neutrino mass, an expression of the orthogonal matrix is  	
\dis{
R =  \left(
\begin{array}{ccc}
		1 & 0          & 0           \\
		0 & \cos\omega & -\sin\omega \\
		0 & \sin\omega & \cos\omega  \\
	\end{array}
\right) ,
\label{Omega}
}
with $\omega$ being a complex parameter.
If the imaginary values for the complex orthogonal matrix are large, 
the components of Yukawa couplings (\ref{eq:NuYukawa}) and mixings (\ref{Theta}) can be enhanced as
\begin{equation}
 |\Theta|^2 = \left|(m^\mathrm{diag}_{\nu}){}^{1/2} R(M_{\nu_R})^{-1/2} \right|^2
 \sim \frac{m_{\nu}}{M_{\nu_R}} \exp( 2\mathrm{Im}\omega ),
	\label{Theta2_Omega}
\end{equation}
which can be much larger than Eq.~(\ref{Theta2}).

By substituting the decomposition in Eqs.~(\ref{Eq:B}), (\ref{Eq:nuL}) and (\ref{Eq:nuR}) into the Lagrangian, we obtain the interaction of the mass eigenstate neutrinos
\begin{align}
	\mathcal{L}_\mathrm{CC} = & \frac{1}{\sqrt{2}}g_2\overline{e}_{\alpha} W^-_{\mu} \gamma^{\mu}
	( U_{\alpha i } P_L \nu_i + \Theta_{\alpha i } P_L \nu_{si}^c ) + \mathrm{H.c.} ,                                                                                                                                                                                                   \\
	\mathcal{L}_\mathrm{NC} = & \frac{1}{2} \sqrt{g_2^2+g_1^2} Z_{\mu}( \bar{\nu}_i  \gamma^{\mu} P_L \nu_i + \overline{\nu^c_{sj}} (\Theta^{\dagger}U)_{ji} \gamma^{\mu} P_L \nu_i
	+  \bar{\nu}_i  \gamma^{\mu} P_L (U^{\dagger}\Theta)_{ij} \nu^c_{sj} +  \overline{\nu^c_{si}} \gamma^{\mu} P_L (\Theta^{\dagger}\Theta)_{ij} \nu^c_{sj} ) ,                                                                                                                         \\
	\mathcal{L}_\mathrm{DI} = & (c_W F_{\mu\nu}-s_W Z_{\mu\nu}) \left\{  \overline{\nu_i} (U^{\dagger}\Theta)_{ik}C_{kl}(\Theta^T U^*)_{lj}[\gamma^{\mu},\gamma^{\nu}] P_R \nu^c_j  \right. \nonumber                                                                                   \\
	                          & \left. -  C_{ij} \overline{\nu_k} (U^{\dagger}\Theta)_{ki}[\gamma^{\mu},\gamma^{\nu}] P_R \nu_{sj}
	-  C_{ij} \overline{\nu_{si}^c}[\gamma^{\mu},\gamma^{\nu}] P_R (\Theta^T U^*)_{jl}\nu^c_l
	+ C_{ij} \overline{\nu_{si}^c}[\gamma^{\mu},\gamma^{\nu}] P_R  \nu_{sj}  \right\}  \nonumber                                                                                                                                                                                        \\
	                          & + (c_W F_{\mu\nu}-s_W Z_{\mu\nu}) \left\{ - \overline{\nu_i^c} (U^T\Theta^*)_{ik}C^{\dagger}_{kl}(\Theta^{\dagger} U)_{lj}[\gamma^{\mu},\gamma^{\nu}] P_L \nu_j  \right. \nonumber                                                                      \\
	                          & \left.
	+\overline{\nu_i^c} (U^T\Theta^*)_{ik}C^{\dagger}_{kj}[\gamma^{\mu},\gamma^{\nu}] P_L \nu^c_{sj}
	+ \overline{\nu_{si}} C^{\dagger}_{ik}(\Theta^{\dagger} U)_{kj}[\gamma^{\mu},\gamma^{\nu}] P_L \nu_j
	- C^{\dagger}_{ij} \overline{\nu_{si}}[\gamma^{\mu},\gamma^{\nu}] P_L \nu_{sj}^c  \right\} \nonumber                                                                                                                                                                                \\
	=                         & (c_W F_{\mu\nu}-s_W Z_{\mu\nu}) \left( \overline{\nu_i} (C_V^{\nu\nu}+C_A^{\nu\nu}\gamma_5)_{ij} [\gamma^{\mu},\gamma^{\nu}]\nu_j + \overline{\nu_i} (C_V^{\nu\nu_s}+C_A^{\nu\nu_s}\gamma_5)_{ij}[\gamma^{\mu},\gamma^{\nu}] \nu_{sj} \right. \nonumber \\
	                          & \left.
	+ \overline{\nu_{si}}(C_V^{\nu_s\nu}+C_A^{\nu_s\nu}\gamma_5)_{ij}[\gamma^{\mu},\gamma^{\nu}] \nu_l
	+ \overline{\nu_{si}}(C_V^{\nu_s\nu_s}+C_A^{\nu_s\nu_s}\gamma_5)_{ij}[\gamma^{\mu},\gamma^{\nu}] \nu_{sj}  \right) .
	\label{LM}
\end{align}
In the last line in Eq.~(\ref{LM}), we used Majorana nature $\nu_i=\nu_i^c$ and $\nu_{sj}=\nu_{sj}^c$, and the corresponding vector and axial-vector couplings are defined as
\dis{
(C_V^{\nu\nu} +C_A^{\nu\nu}\gamma_5)_{ij}=& (U^{\dagger}\Theta)_{ik}C_{kl}(\Theta^T U^*)_{lj} P_L - (U^T\Theta^*)_{ik}C^{\dagger}_{kl}(\Theta^{\dagger} U)_{lj} P_R,\\
(C_V^{\nu\nu_s}+C_A^{\nu\nu_s}\gamma_5)_{ij}=& - (U^{\dagger}\Theta)_{ik}C_{kj} P_L  +(U^T\Theta^*)_{ik}C^{\dagger}_{kj} P_R,\\
(C_V^{\nu_s\nu}+C_A^{\nu_s\nu}\gamma_5)_{ij}=&  - C_{ik}(\Theta^T U^*)_{kj} P_R + C^{\dagger}_{ik}(\Theta^{\dagger} U)_{kj} P_L,\\
(C_V^{\nu_s\nu_s}+C_A^{\nu_s\nu_s}\gamma_5)_{ij} =& C_{ij}P_R - C^\dagger_{ij} P_L,
\label{CVA}
}
with the chiral projection operator
\dis{
	P_{L,R} = \frac{1\mp\gamma_5}{2}.
}

The dipole interaction operator $\overline{\nu_i} [\gamma^{\mu},\gamma^{\nu}]\nu_{sj} F_{\mu\nu}$ can be induced
also from the
Dirac dipole operator, $\overline{\nu}_L [\gamma^{\mu},\gamma^{\nu}]\nu_R B_{\mu\nu}$.
However, this is actually a dimension $6$ operator with~\cite{Bell:2008fm,Duarte:2016miz}
\begin{equation}
	\frac{1}{\Lambda_6^2}\overline{L}\widetilde{\Phi} [\gamma^{\mu},\gamma^{\nu}]\nu_R B_{\mu\nu} \rightarrow \frac{v}{\Lambda_6^2} \overline{\nu} [\gamma^{\mu},\gamma^{\nu}]\nu_s F_{\mu\nu} ,
\end{equation}
while the operator $\overline{\nu_{Ri}^c}[\gamma^{\mu},\gamma^{\nu}] \nu_{Rj} B_{\mu\nu}$ is dimension five.
Thus, as usual if we assume that the cut-off scale is common for all higher dimensional operators, then
the operator with  lower dimension must be more important. In this respect, we do not consider dimension-$6$ operators in the rest of this paper.

\section{Lightest sterile neutrino as dark matter}
\label{sec:dm}

The lightest sterile neutrino can be a good candidate for DM, if it is stable enough and the relic density is consistent with that for DM.
In this section, we study the parameter space in our model for the lightest sterile neutrino to be DM by examining the lifetime and its production in the early Universe. We find that the lightest sterile neutrino with the mass of around MeV and the dimension-five dipole interaction suppressed by $\Lambda_5 \gtrsim 10^{15}$ GeV can be a good candidate for dark matter, and the right amount for DM can be produced through the thermal production in the early Universe. 
Though we denote the mass eigenstate of sterile neutrinos $\nu_{si}$ in the previous section,
to distinguish between the lightest DM sterile neutrino and the heavier sterile neutrinos, we use $\nu_s$ for only the lightest sterile neutrino $\nu_{s1}$ and $\nu_{sh}$ for the heavier sterile neutrinos with $h$ running from $2$ to $3$ in the rest of this paper.

\subsection{Stability of sterile neutrino DM}
\label{subsec:DMLifetime}

Even in the case of vanishing $y_{\nu\alpha 1}$, a $\nu_{s}$ can decay into a photon and an active neutrino $\nu_i$ through the dipole interaction and the mixing between the heavier sterile neutrinos with active neutrinos, $\nu_s \rightarrow \nu \gamma$. 
To avoid the constraints from the monochromatic photon observation, it is required that the lifetime $\tau> 10^{28}$ second, which is much longer than the age of the present Universe.  The decay rate of $\nu_s \rightarrow \nu \gamma$  is given by
\dis{
\Gamma(\nu_s \rightarrow \nu \gamma) \simeq &
\frac{1}{2\pi } c_W^2 \sum_{i=1}^3[ |C^{\nu\nu_s}_{V h1}|^2+| C^{\nu\nu_s}_{A h 1}|^2 ] m_{\nu_{s}}^3  \\
\sim & \frac{1}{10^{28} \sec} \bfrac{10^{15}\gev}{\Lambda_5}^2\bfrac{|\Theta|}{10^{-6}}^2 \bfrac{m_{\nu_s}}{1\mev}^3,
\label{DMtau}
}
where, $i$ is the index of the mass eigenstate of light active neutrinos, and $\Theta$ is the non-vanishing mixing between active and heavier sterile neutrinos ($\nu_{s2}$ and $\nu_{s3}$).
In the second line, we used the relation in \eq{CVA} by dropping the constants of the order of unity $c_{ij}$ and
\begin{equation}
	C_{jk} = \mathcal{O}\left(\frac{1}{\Lambda_5}\right),
\end{equation}
up to complex phases. We find that the preferred parameters are  $\Lambda_5 \gtrsim 10^{15}$ GeV for $m_{\nu_s}\simeq 1$ MeV and $|\Theta| \simeq 10^{-6}$ from \eq{DMtau}.

\subsection{Relic density of sterile neutrino DM}
\label{relicdensity}

The sterile neutrino DM can be produced in the early Universe through the scatterings of the thermal particles (thermal production) and decay of the decoupled heavy unstable particles (non-thermal production). First, we provide the general formula for the production of DM in this subsection and details of the thermal and non-thermal production of the lightest sterile neutrino DM are presented in the~\ref{subsec:TPDMProduction} and~\ref{subsec:NTPDMProduction}, respectively.

The Boltzmann equation for the number density of heavier sterile neutrinos $n_{\nu_{sh}} (h=2,3)$ are described by
\dis{
	\frac{d n_{\nu_{sh}}}{dt} + 3 H n_{\nu_{sh}} = & \langle\sigma v (a b \rightarrow \nu_{sh} X)\rangle (n_a n_b  - n_{\nu sh} n_X )
	-\Gamma(\nu_{sh} \rightarrow \mathrm{all})n_{\nu_{sh}}\label{Bolt_Heavy} 
} 
for $h=2$ and $3$, where $n_a$ are the number density of $a$ particle, $\sigma v(a b \rightarrow \nu X)$ is the scattering cross section times the relative velocity for a process from initial particles $a$ and $b$ to $\nu$ and another by-product $X$, and $\Gamma$ is the decay rate of the corresponding mode.
Similarly, the Boltzmann equation for the number density of the sterile neutrino DM $\nu_{s}$ is given by  
\dis{	
	\frac{d n_{\nu_{s}}}{dt} + 3 H n_{\nu_{s}} =   & \langle\sigma v (a b \rightarrow \nu_{s} X)\rangle n_a n_b + \Gamma(\nu_{sh} \rightarrow \gamma\nu_{s})n_{\nu_{s h}}.
	\label{Bolt}
}
The last term is the production from the decay of the heavier sterile neutrinos $\nu_{sh}$ and we ignored the decay of $\nu_s$ here, since it is negligibly small.
The Hubble parameter $H$ in the radiation dominated Universe is given by
\begin{equation}
	3 M_P^2 H^2 = \rho\simeq  \frac{\pi^2 g_*}{30}T^4,
\end{equation}
where $\rho$ is the total energy density of radiation, $T$ is its temperature, 
 $g_*$ is the total relativistic degrees of freedom,
 and $M_P$ is the reduced Planck mass.

The scattering term in the RHS in Eqs.~(\ref{Bolt_Heavy}) and (\ref{Bolt}) is given by~\cite{Gondolo:1990dk}
\begin{align}
	\langle\sigma v\rangle n_i n_j & =
	\frac{T}{32\pi^4}\sum_{i, j}\int^{\infty}_{(m_i+m_j)^2} ds g_i g_j p_{ij} 4E_{i}E_j \sigma v K_1\left(\frac{\sqrt{s}}{T}\right) ,
\end{align}
where
\begin{align}
4E_{i}E_j \sigma v 
\equiv \prod_f \int \frac{d^3 p_f}{(2\pi)^3}\frac{1}{2E_f}\overline{|\mathcal{M}|^2}(2\pi)^4 \delta^{(4)}(p_i+p_j-\sum p_f) ,
\end{align}
with $f$ being indices for final states, and
\begin{align}
	p_{ij} & \equiv \frac{(s-(m_i+m_j)^2)^{1/2}(s-(m_i-m_j)^2)^{1/2}}{2\sqrt{s}}.
\end{align}
Here, $K_1(z)$ is the modified Bessel function of the first kind with
$g_i$,  $m_i$, and $E_i$ being the internal degrees of freedom, and the mass and energy of $i$ particle, respectively. The Mandelstam variable  $s = (p_i + p_j)^2$ is defined with the incoming 4-momenta of $p_i$ and $p_j$.

For heavier sterile neutrinos, the dominant production comes from the thermal scattering through Yukawa interactions, dominantly through the Higgs boson mediation $q\nu\rightarrow q \nu_{sh}$ (t-channel) and $\bar{q}q\rightarrow \bar{\nu}\nu_{sh}$ (s-channel) for $h=2,3$. Here, any SM fermion can be $q$, however, in practice, (top) quarks are dominant.
Heavier sterile neutrinos also decay into the active neutrinos and also into the lightest sterile neutrino. 
On the other hand, the lightest sterile neutrinos with vanishing Yukawa couplings are produced through the dipole interaction from scattering processes (thermal production, TP) dominantly
$f\bar{f} \rightarrow  \gamma, Z \rightarrow \nu_{s}\nu_{s2}$ or $\nu_s\nu_{s3}$ (s-channel)\footnote{There are other production modes $\gamma W^\pm \, \mathrm{or}\, Z W^\pm \rightarrow \nu_{s2}  \rightarrow \nu_{s} f^{\pm}$(t-channel) and interchange of $W^\pm$ and $f^\pm$. However, those are sub-dominant due to the suppression by both the mixing $\Theta$ and $\Lambda_5$.},
and  from the decay (non-thermal production, NTP) of decoupled heavier sterile neutrinos which such as
$\nu_{s2} (\nu_{s3}) \rightarrow \nu_{s} \gamma$.
Since the interaction of the sterile neutrino DM is too small, their production mechanism is the same as that of gravitino or axino which have very weak interactions~\cite{Baer:2014eja}\footnote{The massive particle with very weak interactions has been called as super-WIMP~\cite{Feng:2003xh}, E-WIMP~\cite{Choi:2005vq}, or FIMP~\cite{Hall:2009bx} }. In fact, operators in Eq.~(\ref{LM}) has the same Lorentz structure to those for axino, namely, the axino-gaugino-gauge field strength vertex~\cite{Choi:1999xm,Covi:1999ty,Covi:2001nw,Choi:2011yf,Choi:2018lxt}. We consider both production mechanisms in the following subsections one by one.

\subsection{Thermal production of sterile neutrino DM}
\label{subsec:TPDMProduction}

The lightest sterile neutrino DM can be  produced directly from the particles in the thermal equilibrium.
Since $\nu_s$ is weakly interacting with the other particles and always in the out of equilibrium, the abundance $Y \equiv n/s$ defined as the ratio of the number density to the entropy density is estimated as
\begin{align}
	Y_{\nu_s}^\mathrm{TP} & = \int_{T_0}^{T_R} \frac{\langle\sigma v (i j \rightarrow \nu_{s1} X)\rangle n_i n_j   + \Gamma(\nu_{sh} \rightarrow \gamma\nu_{s}) n_{\nu_{sh}}}{s T H}dT ,
	\label{Eq:Y0}
\end{align}
where $s = \frac{2\pi^2 g_{*S}}{45}T^3$ is the entropy density with $g_{*S}$ being the total relativistic degrees of freedom for the entropy, $T_0$ is the present temparature and $T_R$ is the reheating temperature after inflation.
Once $Y_{\nu_s}^\mathrm{TP}$ is obtained,  the present relic density is given by
\dis{
\Omega_{\nus}^\mathrm{TP} h^2 &= \frac{m_{\nu_{s}}}{\rho_\mathrm{crit}/s_0}Y_{\nu_s}^\mathrm{TP}\simeq 0.28 \bfrac{m_{\nu_{s}}}{1\mev}\bfrac{Y_{\nu_s}^\mathrm{TP}}{10^{-6}},
\label{Eq:Omegah2_Int}
}
where we used $(\rho_\mathrm{crit}/s_0)^{-1} = 2.8 \times 10^8/\mathrm{GeV}$ with the present entropy density $s_0= \frac{2\pi^2}{45}\times 3.91\times T_0^3$ and the critical density $\rho_\mathrm{crit}=3\Mp^2 H_0^2$ with $H_0$ being the present Hubble parameter.

\begin{figure}[htbp]
\centering
\includegraphics[width=0.7\textwidth]{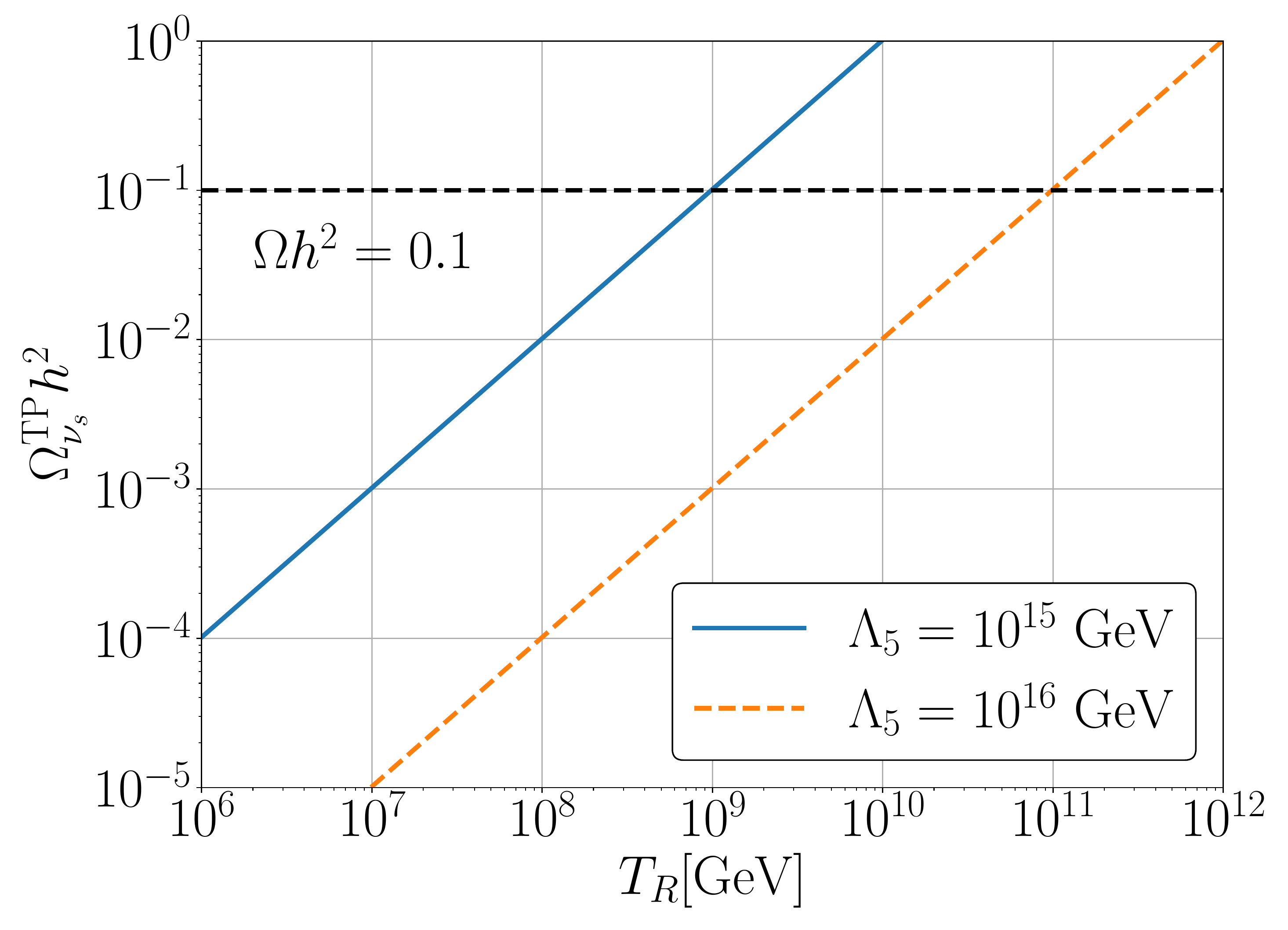}
\caption{The plot of $\Omega^{\mathrm{TP}}_{\nu_s} h^2$ vs $T_R$ with $m_{\nu_s}= 1\mev$ for $\Lambda_5=10^{15}\gev$~(Blue solid) and $10^{16}\gev$~(Orange dashed).}
\label{OmegaTP_TR}
\end{figure}

\begin{figure}[htbp]
\centering
			\includegraphics[width=0.7\textwidth]{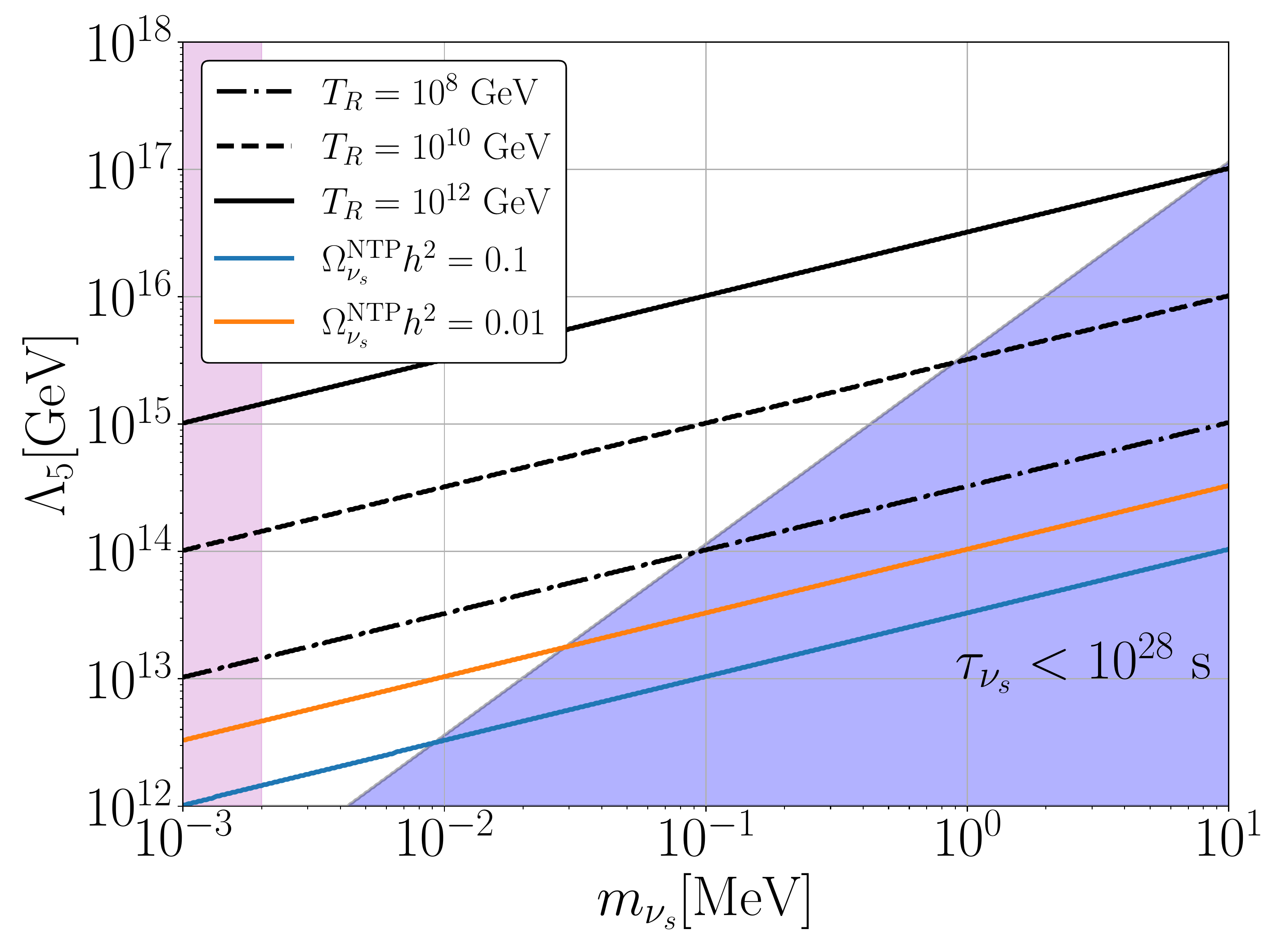}
\caption{The contours on the plane of ($m_{\nu_s}, \Lambda_5$) which give the correct relic density for the lightest sterile neutrino DM from TP, for the reheating temperature $T_R =10^8,10^{10},10^{12}\gev$ with solid, dotted, dot-dashed lines, respectively. Blue, and orange  lines are the amount produced by the nonthermal production as discussed in Sec.~\ref{subsubsec:NTPDMProduction}. The mauve shaded regions are disfavored due to a shorter lifetime of DM than $10^{28}\sec$. The light magenta region, $m_{\nu_s} \lesssim 2\, \kev$, is disfavored from the constraint of the structure formation. Here, we fixed the heavier sterile neutrino masses as $(m_{s2}, m_{s3})= (1 \,\mathrm{GeV}, 10 \,\mathrm{GeV})$.}
	\label{Total}
\end{figure}

At high temperature before the electroweak symmetry breaking,
the most dominant production modes for the lightest sterile neutrino are
the scattering $f \nu_{s2}$ or $f\nu_{s3} \rightarrow f \nu_{s}$ via t-channel
and $f\bar{f} \rightarrow \nu_{s}\nu_{s2}$ or $\nu_{s}\nu_{s3}$ via s-channel
mediated by $B$ boson through the dipole interaction.
In calculation, we introduced the thermal mass for the $B$ boson $m_B\sim g_Y T$ to regulate the divergence in the $t$-channel.
We give the explicit expression of the spin averaged amplitude squared in Appendix.
By substituting thoes into Eqs.~(\ref{Eq:Y0}) and (\ref{Eq:Omegah2_Int}), we obtain
\begin{align}
	Y_{\nu_s}^\mathrm{TP}                            
	\simeq  4.2 \times 10^{-7} \bfrac{10^{16}\gev}{\Lambda_5}^2 \bfrac{T_R}{10^{11}\gev},
\end{align}
and 
\dis{
	\Omega_{\nus}^\mathrm{TP} h^2 & \simeq 0.1 \bfrac{\ms}{1\mev}\bfrac{10^{16}\gev}{\Lambda_5}^2 \bfrac{T_R}{10^{11}\gev}.
	\label{Oh2TP}
}
Note that here the abundance is proportional to the reheating temperature $T_R$ due to the non-renomalizable interaction. In this case, the production from decay of thermal particles are subdominant and can be ignored~\cite{Covi:2001nw}.
We consider the reheating temperature to be less than $\Lambda_5$, otherwise we might have to consider the full UV theory for the temperature above the cutoff scale $\Lambda_5$.
In Fig.~\ref{OmegaTP_TR}, we show the final abundance as a function of $T_R$ for two choices of $\Lambda_5 = 10^{15}$ GeV and $10^{16}$ GeV. In Fig.~\ref{Total}, the contours of $\Omega^{\mathrm{TP}} h^2 = 0.1$ for several $T_R$ are shown with the lifetime constraint with \eq{DMtau} in the ($\ms$, $\Lambda_5$) plane. 

Here, we note that the lightest sterile neutrino DM, $\nu_{s}$, cannot be produced through neutrino oscillation at all,
since the Yukawa coupling $y_{\nu\alpha 1}$ are almost vanishing, thus neither Dodelson-Widrow~\cite{Dodelson:1993je} nor Shi-Fuller~\cite{Shi:1998km} mechanism works.


\subsection{Non-thermal production of sterile neutrino DM}
\label{subsec:NTPDMProduction}

In the hot and dense early Universe, the sterile neutrinos are produced from the thermal particles. After the heavier sterile neutrinos decoupled from thermal bath, those decay into the SM particles as well as the lightest sterile neutrino DM. This contribution to the sterile neutrino DM is called non-thermal production.
For this, we first estimate the relic abundance of the heavier sterile neutrinos at the decoupling time, and then calculate the branching ratio of the decay into the lightest sterile neutrino DM. 

\subsubsection{Production of heavier sterile neutrinos}

The heavier sterile neutrinos $\nu_{sh}$ are produced by scatterings through Yukawa interaction.
The dominant production modes are $q \nu \rightarrow q \nu_{sh}$ via t-channel Higgs boson exchange
 and 
$\bar{q} q \rightarrow \bar{\nu} \nu_{sh} $ via s-channel Higgs boson exchange process.
We express those amplitudes in the Appendix.
The scattering cross section of each mode is given as
\dis{
	\langle\sigma v \rangle \simeq \frac{N_c y_\mathrm{f}^2 y_\nu^2}{128 \pi T^2},
\label{eq:sigv_with_top}
}
for large $T \gg m_t,m_h$, $M_{\nu_R}$.
 Here, $N_c$ is the color factor, $y_\mathrm{f}$ is Yukawa coupling of the SM fermions, $m_t$ and $m_h$ are masses of the top quark and the SM Higgs boson, respectively. 
For the scattering with a top quark, 
the equilibrium condition is expressed as 
\dis{
	n \VEV{\sigma(q \nu \leftrightarrow q \nu_{sh}) v} > H
\label{eq:equilib_top}
}
with $n=\frac{\zeta(3) g_\nu}{\pi^2} T^3$, $\zeta(3)\simeq 1.202$ being the Riemann zeta function of $3$.
By recasting the condition \eq{eq:equilib_top} with Eq.~(\ref{eq:sigv_with_top}), we find that 
the thermal equilibrium is attained for temperature
\begin{align}
	T < 1\tev   \bfrac{y_{\nu}}{10^{-6}}^2 .
   \label{eq:thermal_eq_top}
\end{align}
Since we find 
\begin{align}
	y_{\nu} \sim 10^{-6}\left(\frac{M_{\nu_R}}{100 \, \mathrm{GeV}}\right)^{1/2} R ,\nonumber
\end{align}
from Eq.~(\ref{eq:NuYukawa}), this condition \eq{eq:thermal_eq_top} is satisfied for $M_{\nu_R} \gtrsim 1 \tev$ with $R =I$ and
 for $M_{\nu_R} \ll 1 \tev$ with a nontrivial $R$.
After the electroweak symmetry breaking, top quarks decay and disappear from the thermal bath so that  
the scatterings are suppressed and the heavier sterile neutrinos become decoupled. 
Since sterile neutrinos are relativistic at that moment, the abundance before the decay of heavier sterile neutrinos is given by
\dis{
	Y_{\nu_{s_h}}^{\mathrm{ dec}} (T) \sim \frac{1}{g_{*S}(T_{\mathrm{dec}})} \simeq 10^{-2},\quad {\rm for} \quad h=2,3
	\label{Y2eq}
}
where $T_{\rm dec}$ is the decoupling temperature. 

The heavier sterile neutrinos $\nu_{sh} (h=2,3)$ can also be produced though oscillations from the active neutrinos, namely the Dodelson-Widrow mechanism~\cite{Dodelson:1993je}.
If those were stable, the present abundance could be expressed by~\cite{Dodelson:1993je,Boyarsky:2018tvu}
\begin{align}
	\Omega_{\nu_{sh}} h^2 & \simeq
	0.1 \left(\frac{|\Theta_{\alpha h}|}{1.57 \times 10^{-5}}\right)^2 \left(\frac{m_{\nu_{sh}}}{10~\mathrm{keV}}\right)^2
	\simeq 10^7 \left(\frac{|\Theta_{\alpha h}|}{1.57 \times 10^{-6}}\right)^2 \left(\frac{m_{\nu_{sh}}}{1~\mathrm{GeV}}\right)^2 ,
\end{align}
for $h=2,3$,
which is rewritten as
\dis{
	Y_{\nu_{sh}} \simeq 10^{-2} \left(\frac{|\Theta_{\alpha h}|}{ 10^{-6}}\right)^2\left(\frac{m_{sh}}{1~\mathrm{GeV}}\right), \quad {\rm for} \quad h=2,3.
}
This can be comparable to \eq{Y2eq} of the thermal abundance. For $|\Theta_{\alpha h}| > 10^{-6} (1 \, \mathrm{ GeV}/m_{sh})^{1/2}$, heavier sterile neutrinos could be thermalized by oscillation. 
In the following, we assume that the heavier sterile neutrinos are in the thermal equilibrium, because thermalizaion by either scattering or oscillation is possible for wide parameters of our interest.

\subsubsection{Decay of heavier sterile neutrinos}
\label{subsubsec:NTPDMProduction}

\begin{figure}[!t]
	\begin{center}
		\begin{tabular}{cc}
			\includegraphics[width=0.5\textwidth]{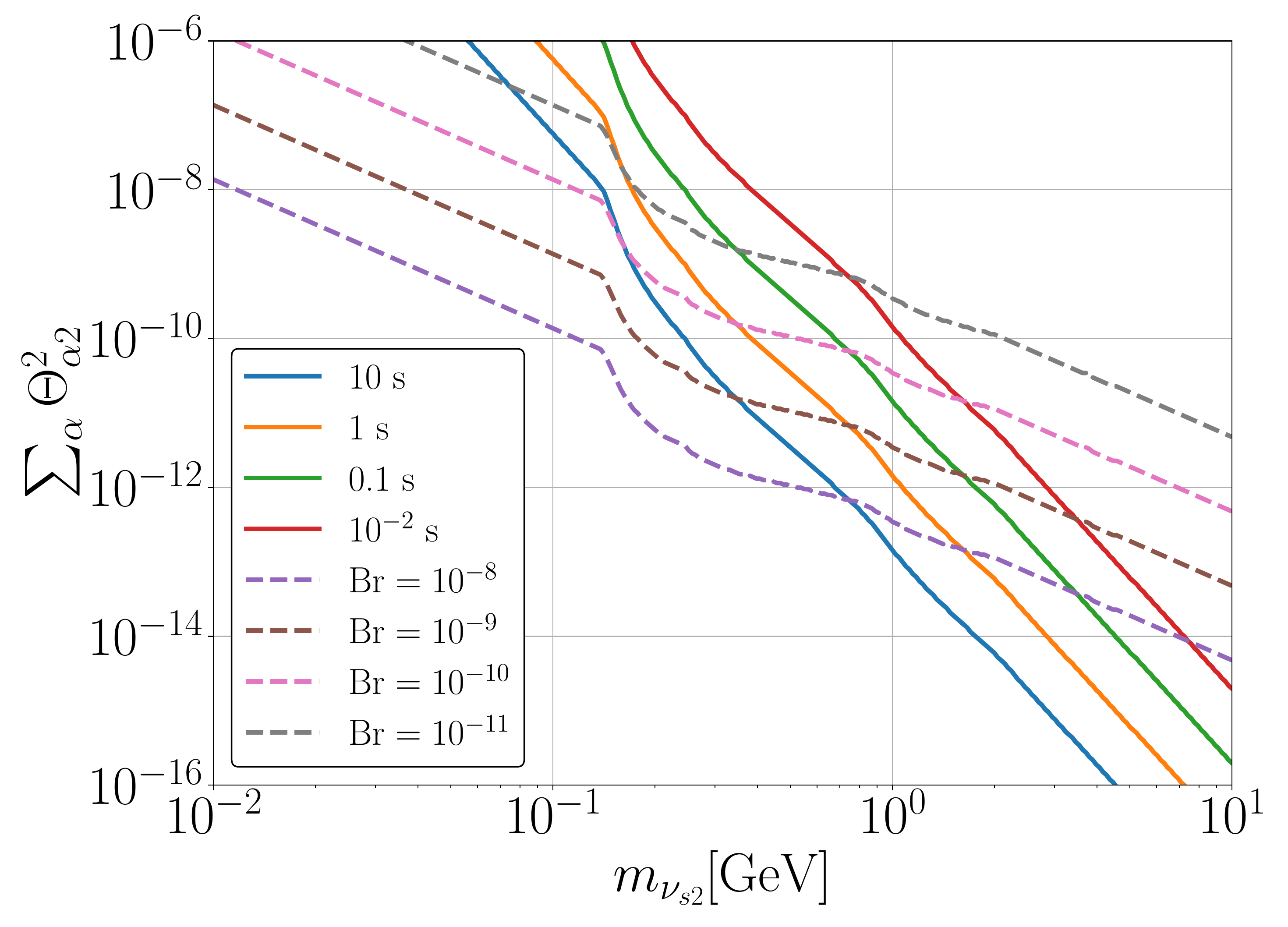}
			 &
			\includegraphics[width=0.5\textwidth]{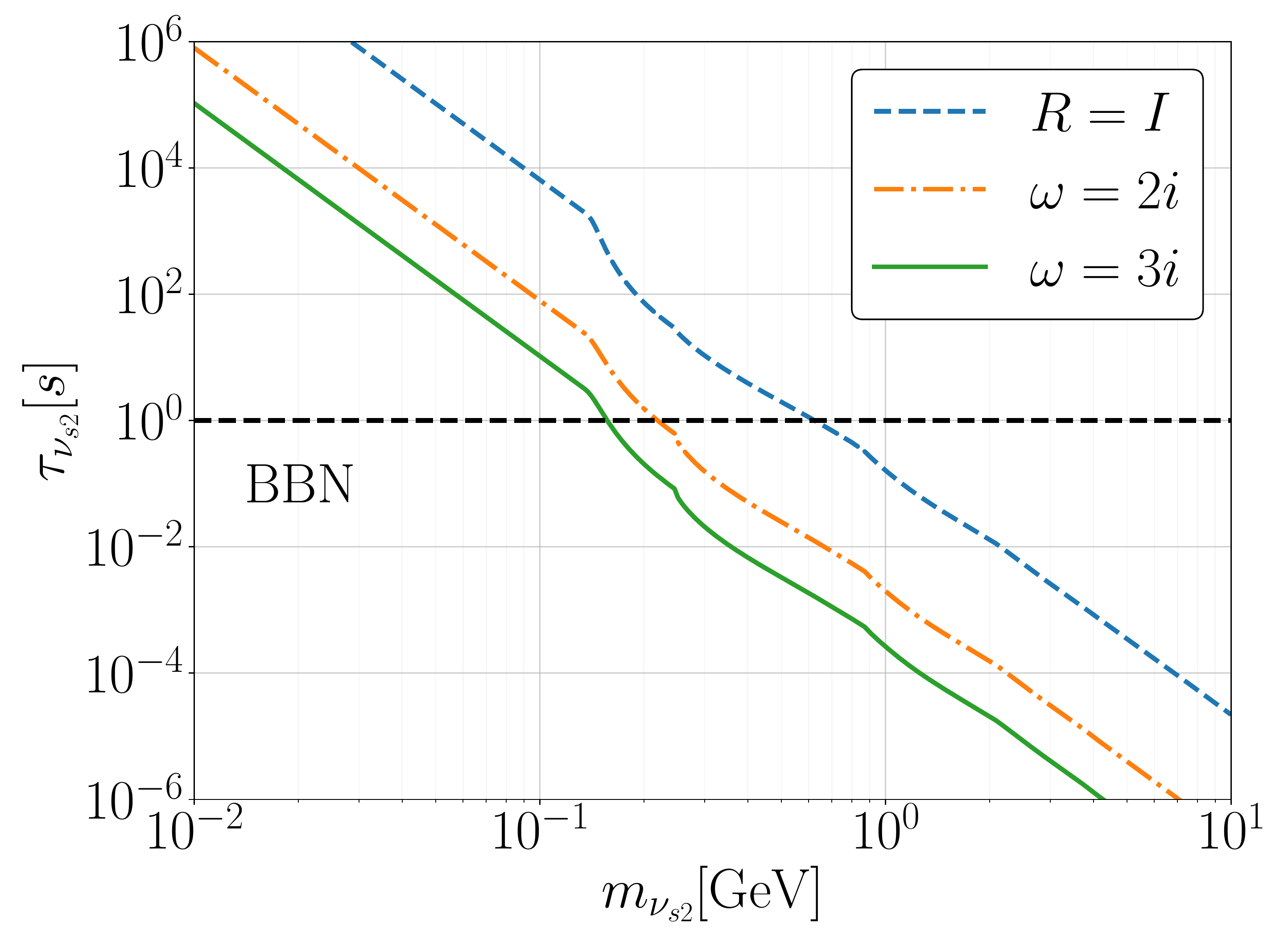}
		\end{tabular}
	\end{center}
	\caption{\textit{Left}: Contours of the lifetime of the heavier sterile neutrino (solid curves) and the decay branching ratio $\mathrm{Br}(\nu_{s2}\rightarrow \gamma\nu_s)$ (dashed curves) on the plane of $m_{\nu_{s2}}$ vs $\sum_{\alpha}|\Theta_{\alpha j}|^2$ for fixed $m_{\nu_{s3}}=10$ GeV. \textit{Right}: Lifetime of heavier sterile neutrino for different orthogonal matrix with $R = I$, $\omega = 2i$ and $\omega = 3i$.}
	\label{Lifetime}
\end{figure}

The heavier sterile neutrinos can decay in the early Universe due to the Yukawa interaction as well as the dipole term.
The decay modes due to Yukawa interaction include
$\nu_{sh} \rightarrow 3\nu$, with other leptonic decay modes such as $\ell^-\ell^+\nu$~\cite{Asaka:2012hc}, $\nu_{sh} \rightarrow \nu\gamma$~\cite{DeRujula:1980mgi,Pal:1981rm}, and the modes with mesons.
The partial decay rate can be found in Refs.~\cite{Atre:2009rg,Ballett:2016opr} for the sterile neutrino lighter than the $W$-boson.
Among them, the decay rate of the dominant decay mode is
\dis{
	\Gamma(\nu_{sh} \rightarrow 3\nu )&= \sum_{\alpha, h} \Gamma(\nu_{sh} \rightarrow \nu_\alpha \nu_i\bar{\nu}_i) = \frac{G_F^2 m_{\nu_{sh}}^5}{96\pi^3} \sum_{\alpha}|\Theta_{\alpha h}|^2,
}
which estimates the lifetime of the heavier sterile neutrinos. In the left window of Fig.~\ref{Lifetime}, we show the contour of the lifetime of the heavier sterile neutrinos in the plane of $(m_{\nu_{sh}}, \sum_\alpha |\Theta_{\alpha j}|^2)$. The solid line corresponds to the lifetime ($10,1,0.1,0.01$) second, respectively  from left to right. In the right window,  we show the lifetime of the heavier sterile neutrino for different $R$ of $R=I$ and $R$ with $\omega=2i$, and $3i$ as a function of $m_{\nu_{s2}}$ with the fixed $m_{\nu_{s3}}=10$ GeV. For each $R$, the mixings are $\sum_\alpha |\Theta_{\alpha h}|^2=8\times 10^{-12},~8\times 10^{-10},~6\times 10^{-9} $ for $m_{\nu_{s2}}=1\gev$, respectively.

The late decay of heavier sterile neutrinos may disrupt the standard process of BBN and recombination~\cite{Dolgov:2000jw,Ruchayskiy:2012si,Vincent:2014rja}. To avoid these problems, it is generally required that the heavy particles decay before around $1$ second of the age of the Universe. As is seen in Fig.~\ref{Lifetime}, this is satisfied if the mass of the heavier neutrino is larger than around hundred MeV.

The dipole interaction allows new decay mode $\nu_{sh} \rightarrow \gamma\nu_s$ and contributes to the  nonthermal production of $\nus$~\cite{Nemevsek:2012cd}. The decay width for this process is given by
\begin{align}
	\Gamma(\nush \rightarrow \nus \gamma) & =\int\frac{|\mathbf{p}|}{8 \pi m_{\nu_{sh}}^2}\overline{|\mathcal{M}|^2}\frac{d\Omega}{4 \pi} \nonumber \\
	                                      & = \frac{1}{2\pi } c_W^2 [ (C^{\nu_s\nu_s}_{V h1})^2+( C^{\nu_s\nu_s}_{A h 1})^2 ] m_{\nu_{sh}}^3,
\end{align}
where $C^{\nu_s\nu_s}_{V h1}$ and $C^{\nu_s\nu_s}_{A h 1}$ are defined in \eq{LM} and $\mathbf{p}$ is the three momentum of the final state and $d\Omega$ is integration wih repect to the solid angle.
At high temperature before the $SU(2)_L \times U(1)_Y$ symmetry breaking,
the decay $\nush \rightarrow  \nus B$ has the decay rate
\begin{align}
	\Gamma(\nush \rightarrow \nus B)
	= \frac{1}{2\pi} (C_{h1})^2  m_{\nu_{sh}}^3 .
\end{align}
%

\begin{figure}[!t]
	\begin{center}
		\begin{tabular}{cc}
			\includegraphics[width=0.5\textwidth]{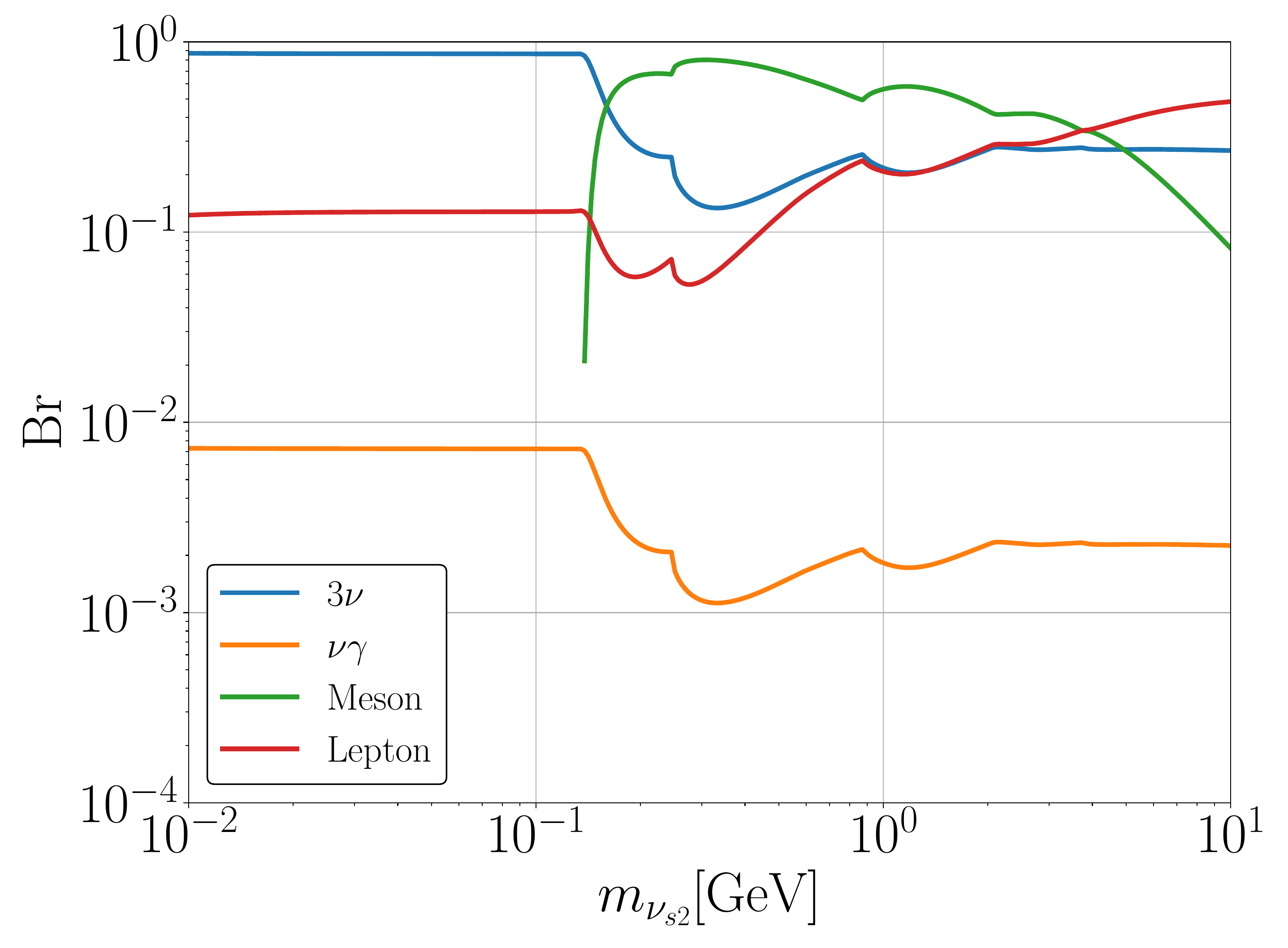}
			 &
			\includegraphics[width=0.5\textwidth]{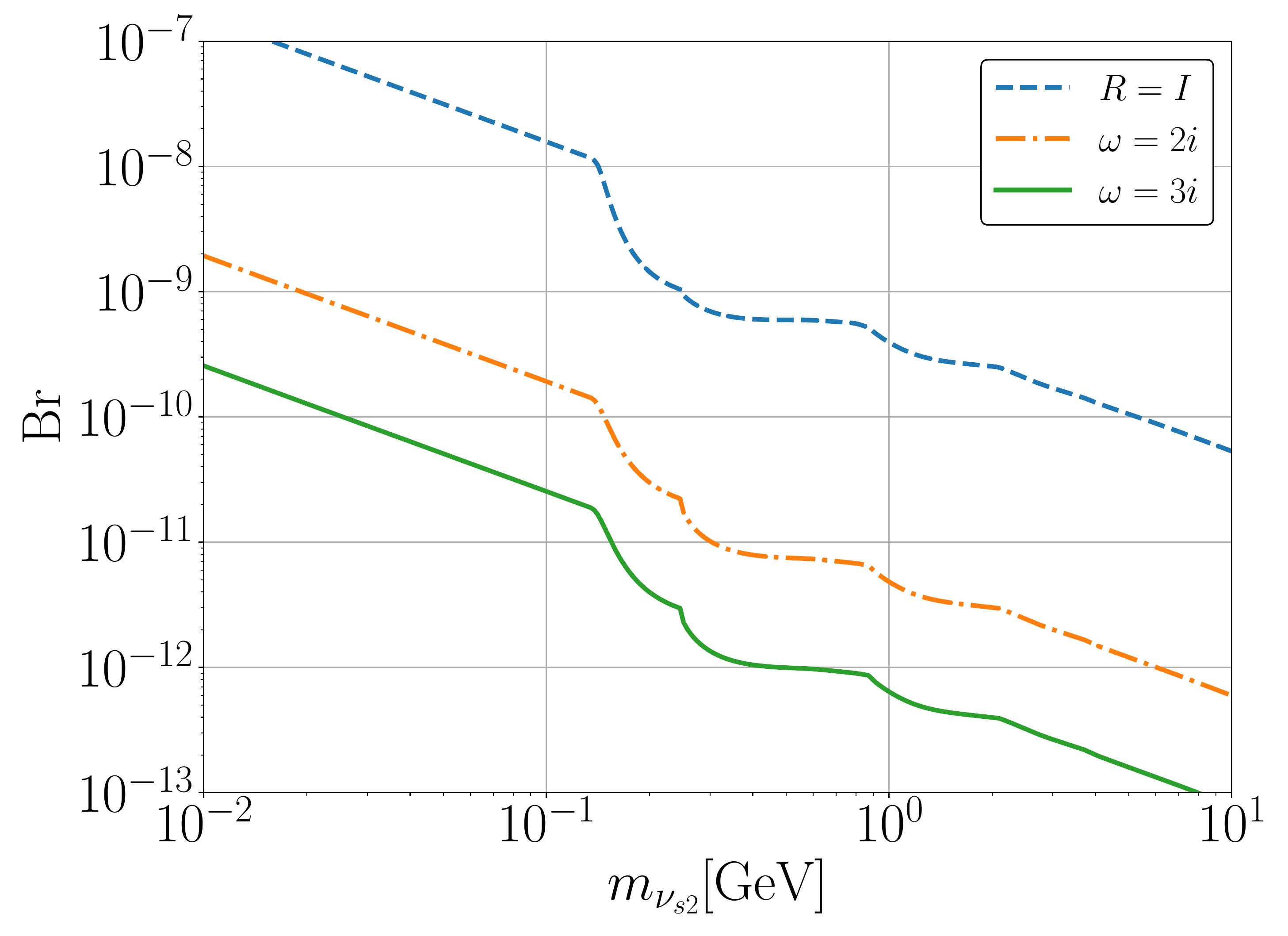}
		\end{tabular}
	\end{center}
	\caption{Branching ratios of the various decay modes of heavier sterile neutrino. \textit{Left}: Branching ratios for the dominant modes, $\nu_{s2} \rightarrow 3\nu$~(blue), $\nu_{s2} \rightarrow \nu\gamma$~(orange), the sum of hadronic modes~(green) and the sum of (charged) leptonic modes~(red) for fixed $m_{\nu_{s3}}=10$ GeV. This is for $R= I$, while difference of the cases with a nonvanishing imaginary $\omega$ is about a factor and is not significant. \textit{Right}: The branching ratio for the decay mode to $\nu_s \gamma$. Here, we used $R= I$~(dashed blue), $R=2i$~(dot-dashed orange), $R=3i$~(solid green), and fixed $\Lambda_5=10^{16}\gev$. }
	\label{BR}
\end{figure}

The abundance of the lightest sterile neutrino DM from the decay of the heavier sterile neutrinos is the fraction of the abundance of them as
\dis{
Y_{\nu_s}^\mathrm{NTP} = {\rm Br}(\nu_{sh} \rightarrow \nu_s  \gamma) \times Y_{\nu_{s_h}}^\mathrm{dec} .
}
The branching ratio is given approximately as
\dis{
	{\rm Br}(\nu_{sh} \rightarrow  \nu_s  \gamma) \simeq \frac{\Gamma(\nu_{sh} \rightarrow \nu_s \gamma ) }{\Gamma(\nu_{s h} \rightarrow 3\nu )}
	\simeq \frac{48\pi^2}{\Lambda_5^2G_F^2 m_{\nu_{sh}}^2 \sum_\alpha|\Theta_{\alpha h}|^2 },
}
which is about $5 \times 10^{-10}$ for $\Lambda_5=10^{16}\gev$, $m_{\nu_{sh}}=1\gev$, and $R=I$.
We display the branching ratio of the major decay modes and the $\nu_{sh} \rightarrow \nu_s \gamma$ mode
 in the heavier sterile neutrino decay in the left panel and the right panel of Fig.~\ref{BR}, respectively. 
The former is proportional to the square of the mixing $\Theta$, 
while the latter is inversely proportional to the square of the cutoff scale $\Lambda_5$.
For $R = I$, the branching ratio ${\rm Br}(\nu_{sh} \rightarrow \nu_s \gamma)$ and,
 as the result,  the abundance $Y_{\nu_s}^\mathrm{NTP}$ is maximized.

The final relic density from the NTP can be written as
\begin{align}
	\Omega_{\nus}^\mathrm{NTP} h^2 = & \frac{ m_{\nu_s} s_0}{\rho_\mathrm{crit}} \sum_{h=2,3} {\rm Br}(\nu_{sh} \rightarrow \gamma \nu_s ) \times Y_{\nu_{s_h}}^{\mathrm{dec}}  \nonumber                                  \\
	\simeq  & 1 \times 10^{-6} \bfrac{\ms}{1\mev} \bfrac{1\gev}{m_{\nu_{sh}}}\bfrac{10^{16}\gev}{\Lambda_5}^2\bfrac{9 \times 10^{-12}}{ \sum_\alpha|\Theta_{\alpha h}|^2}.
\end{align}
Note that the heavier sterile neutrinos are decoupled after $T< m_t$ and their abundance is frozen so that there is no Boltzmann suppression at low temperatures.
As shown in Fig.~\ref{Total},  in the most of the parameter regions, the NTP of the lightest sterile neutrino DM is negligible compared to the thermal production.
This can be understood from the left panel of Fig.~\ref{BR} that shows the strong correlation between the lifetime of a heavier sterile neutrino and its branching ratio into the lightest sterile neutrino. For the parameters where the lifetime of a heavier sterile neutrino is short enough for BBN, the branching ratio into the lightest sterile neutrino cannot be large. 
Although it appears that NTP contributes with the significant fraction of the abundance
 for $\Lambda_5 = \mathcal{O}(10^{12})$ GeV and $m_{\nu_s} < 10$ keV,
 this case is excluded by the constraints from the free-streaming~\cite{Covi:2001nw,Cembranos:2005us,Hisano:2006cj}.
For $m_{\nu_{s2}} = 1$ GeV and the decay temperature of $\nu_{\nu_{s2}}$ of $10$ MeV, 
we obtain  $m_{\nu_s} > \mathcal{O}(100)$ keV by the free streaming constraint that DM must be non-relativistic until the temperature becomes about $\mathcal{O}(1)$ keV.
Those are conflicting. Hence, the NTP contribution cannot be significant to abundance.

\section{Observational Signatures}
\label{sec:test}

As in the usual keV-scale sterile neutrino DM in the $\nu$MSM,
 our sterile neutrino DM with dipole interaction also can be searched by its indirect detection of the decay of DM.
While the dominant decay mode would be $\nu_s \rightarrow 3\nu$,
 the most visible decay mode is $\nu_s \rightarrow \gamma\nu$ with the decay rate \eq{DMtau}.
Since the DM mass of our interest spans from keV to MeV in our model,
the monochromatic X-ray or gamma-ray with the energy $E_\gamma = m_{\nu_s}/2$ is generated. 

One of the possible signature of our DM would be a line-like spectrum in X- or gamma-rays, as in ``$3.5$ keV anomaly''~\cite{Bulbul:2014sua,Boyarsky:2014jta}.
Our model hardly explains the $3.5$ keV anomaly nevertheless, because the parameter sets of the corresponding mass and lifetime are excluded by the free streaming length of the non-thermally produced DM as discussed just above.
Although this signature looks same as that in the $\nu$MSM~\cite{Adhikari:2016bei,Boyarsky:2018tvu},
 an advantage of our model is the fact that cosmological abundance can be explained consistently
 as discussed above.

If kinematically possible, the sterile neutrino DM decays into a pair of electron and positron, and neutrino, $\nu_s \rightarrow e^- e^+ \nu_i$.
The detection of those electrons and positrons could be a signal.
In our model with dipole interaction, the relation between decay rates is predicted as~\cite{Choi:2019pos}
\dis{
\frac{\Gamma(\nu_s \rightarrow e^- e^+ \nu_i)}{\Gamma(\nu_s \rightarrow \gamma \nu_i)}
\simeq \alpha_\mathrm{em} & \left[  \left(1-4\frac{m_e^6}{m_{\nu_s}^6}\right) \ln\left(\sqrt{\frac{m_{\nu_s}^2}{4m_{e}^2}-1}+ \frac{m_{\nu_s}}{2m_{e}} \right) \right. \\
& \qquad \left. - \frac12 \sqrt{1-\frac{4 m_e^2}{m_{\nu_s}^2}}\left( 3-5\frac{4m_{e}^2}{m_{\nu_s}^2} +2 \frac{4m_{e}^4}{m_{\nu_s}^4} \right) \right] , \\
\sim  \alpha_\mathrm{em} & \left( \ln \frac{m_{\nu_s}}{m_e} -\frac32\right),
\label{Eq:DecayRatio} 
}
where the mass of the active neutrinos are neglected and $m_{\nu_s} \gg m_e$  is imposed in the second line.
 Considering the constraints $ \Gamma^{-1}(\nu_s \rightarrow \gamma \nu_i) > 10^{28}\sec$ from monochromatic photon lines,
 our model predicts that  $\Gamma^{-1}(\nu_s \rightarrow e^- e^+ \nu_i) > 2\times10^{29}\sec \sim 2\times10^{30}\sec$ for the DM mass between a few MeV and $10\gev$. These values are much larger than the current lower limit of $\mathcal{O}(10^{26}) \sec$ from the cosmic ray observations~\cite{Cummings:2016pdr,Boudaud:2016mos}.
 In future, if both electron-positron and gamma ray excesses would be discovered with the strength ratio as \eq{Eq:DecayRatio}, it would support our scenario. 
 
If the mass of $\nu_s$ is larger than twice of electron mass $2m_{e}$, the sterile neutrino DM can produce electron and positron through its decay $\nu_s \rightarrow e^- e^+ \nu_i$. The positrons can lose energy after production and form positronium with the background electrons, which decay and contribute to the mohochromatic photons of $511$ keV. In fact, such X-ray line excess from the Galactic Bulge has been reported by the INTEGRAL/SPI~\cite{Knodlseder:2003sv,Jean:2003ci}~\footnote{For its annihilating DM interpretation, see e.g., Refs.~\cite{Boehm:2003bt,Hooper:2008im,Farzan:2020llg}. The decaying DM interpretation proposed e.g., in Refs.~\cite{Hooper:2004qf,Picciotto:2004rp,Khalil:2008kp} is excluded~\cite{Vincent:2012an}.}.
The contribution to the 511 keV line in our model is expected as~\cite{Hooper:2004qf,Picciotto:2004rp}
\dis{
\Phi_{511} \sim 10^{-5} \bfrac{10^{29}\sec}{\Gamma^{-1}(\nu_s \rightarrow e^- e^+ \nu_i)} \bfrac{1\mev}{m_{\nu_s}} \cm^{-2} \sec^{-1} ,
}
which is $0.01$ times smaller compared to the INTEGRAL/SPI $511$ keV line excess and consistent with line gamma searches in dwarf galaxies~\cite{Siegert:2016ijv}.

\section{Conclusion}
\label{sec:concl}

We studied the possibility of the lightest sterile neutrino as dark matter in the presence of the dipole interaction term between the sterile neutrinos.
Sterile neutrino DM with the mass from sub-MeV to MeV scale can be produced thermally and the abundance is proportional to the reheating temperature after inflation and inversely proportional to the square of the cut off scale of the dipole operator.
In other words, if an ultraviolet theory predicts this dipole operator, there is the upper bound on the reheating temperature for DM sterile neutrinos or sterile neutrinos should decay and cannot be a DM candidate.
On the other hand, NTP is severely constrained from the structure formation, because nonthermally produced component are too warm if it constitutes the dominant part.

Here are a few remarks.
We note that the interesting $\Lambda_5$ scale is too large to be constrained by any terrestrial experiment.
A stringent astrophysical constraint would come from stellar cooling. Magill et al reported that the SN bound disappears for $|d| < 10^{-11}\gev^{-1}$ and showed it in Fig.~11 of their paper~\cite{Magill:2018jla}.
In our model, the energy loss rate depends on $\Theta/\Lambda_5$, which is much smaller than $10^{-11}\gev^{-1}$, since the $\nu_s$ coupling to the SM particles are suppressed by both the dipole term and the mixing.
Thus, the $\Lambda_5$ scale of our interest is free from stellar constraints as well.
Throughout our analysis, we have taken heavier sterile neutrino mass to be $O(1)$ GeV and shown its viability.
Thus, in our scenario, the baryon asymmetry in our Universe also could be explained by the mechanism so-called ``baryogenesis via neutrino oscillation''~\cite{Akhmedov:1998qx,Asaka:2005pn}.

\section*{Acknowledgments}
We acknowledge NRF-JSPS Bilateral Open Partnership Joint Research Projects (NRF-2020K2A9A2A08000097).
W.C and K.-Y.C. were supported by the National Research Foundation of Korea (NRF) grant funded by the Korea government (MEST) (NRF-2019R1A2B5B01070181).
This work of O.S. was supported in part by the Japan Society for the Promotion of Science (JSPS) / The Ministry of Education, Culture, Sports, Science and Technology (MEXT) KAKENHI Grants No.~19K03860, No.~19K03865 and No.~21H00060.

\appendix
\section{Amplitude}

In this Appendix, we present the expression of the formulae used in the calculation of the decay of sterile neutrinos and scattering processes involved in the DM production. 

The differential cross section for $2 \rightarrow 2$ scattering with the initial and final momemtums ($ p_1 p_2 \rightarrow p_3 p_4$) in the center of mass (COM) frame is obtained from the scattering matrix element by
\dis{
	\frac{d\sigma }{dt} = \frac{1}{64\pi s} \frac{1}{|{\bf p}_1|^2} \overline{|\mathcal{M}|^2}, 
}
where ${\bf p}_1$ is 3-momentum of one initial particle $p_1$ in the COM frame.
In the massless limit,  the total scattering cross section is obtained by integrating differential cross section as
\dis{
	\sigma =& \int_{-s}^0 \frac{d\sigma }{dt} dt = \frac{1}{16\pi s}  \int \frac{1}{2} |\mathcal{M}|^2 d\cos\theta,
}
where $dt = \frac{s}{2}d\cos\theta$.
Here and in the following, $\theta$ is a scattering angle and
$s$ is the energy-squared in the COM frame.

We give explicit formulas of the spin averaged invariant amplitude squared for the decay and pair annihilation processes
of the RH neutrinos. The relevant couplings can be found in Eqs.~(\ref{LM}) and (\ref{CVA}). 

\subsection{$\nus \rightarrow \nu_i \gamma$}
%
\begin{equation}
	\overline{|\mathcal{M}|^2} = 8 \ms^4  [ (C^{\nu_s\nu_s}_{V i1})^2+( C^{\nu_s\nu_s}_{A i 1})^2 ].
\end{equation}
%
\subsection{$\nusj \rightarrow \nus B $}
%
\begin{equation}
	\overline{|\mathcal{M}|^2} = 4 \left( \msj^2- \ms^2 \right)^2.
\end{equation}
%
\subsection{$f\bar{f} \rightarrow \nu_{s1}\nu_{sj}(\nu_{si} \nu_{sj} )$ via $s$-channel $B$ exchange}
%
\begin{align}
	\overline{|\mathcal{M}|^2} = 4 g_Y^2 Y_f^2 C_{i j}^2 & \left(\frac{ \ms^2 (4 \msj^2 m_f^2 + s (s + 2 t)) - (\ms^4 (2 m_f^2 + s))}{s^2} \right. \nonumber \\
	                                                     & \left. + \frac{- \msj^4 (2 m_f^2 + s) +  \msj^2 s (s + 2 t)}{s^2} 
+ \frac{2 s (m_f^2-t) (-m_f^2+s+t)}{s^2} \right) \nonumber,
\end{align}
In the massless limit of the external particles, this is simplified as
\begin{equation}
	\int\frac{1}{2}
	\overline{|\mathcal{M}|^2} d \cos\theta = \frac{4}{3}s \times N_c (Y_f g_Y)^2C_{1j}^2,
\end{equation}
where $g_Y$ is the $U(1)_Y$ gauge coupling, $Y_f$ is the charge for fermion $f$ and $N_c$ is the color factor.

\subsection{$ f \nu_{sj}  \rightarrow f \nu_{si}  $ via $t$-channel $B$ exchange}
%
\begin{align}
	\overline{|\mathcal{M}|^2} = 4 N_cg_Y^2 Y_f^2 C_{i j}^2 & \left(\frac{-2 m_f^2 (\msi^2 - \msj^2)^2}{(t-m_B)^2} 
+ \frac{t^2 (\msi^2 + \msj^2 + 2 m_f^2 - 2 s)}{(t-m_B)^2} \right.                                         \\
	                                                        & \left. + \frac{- t (-2 \msi^2 s + \msi^4 - 2 \msj^2 s + \msj^4 + 2 (m_f^2 - s)^2)}{(t-m_B)^2} \right) ,\nonumber
\end{align}
To regularize the divergence in the massless limit of $m_B$, we consider the thermal mass of $B$-boson  as $m_B\sim g_Y T$. In the massless limit of the external particles, this is simplified as
\begin{align}
	\int\frac{d \cos\theta}{2}
	\overline{|\mathcal{M}|^2} =
	8 N_c y_f^2 g_Y^2C_{ij}^2  \left( -2s- (s+2m_B^2)  \log\frac{m_B^2}{ s+ m_B^2}\right).
\end{align}
%
\subsection{$f \nu_L \rightarrow f \nu_s $ via $t$-channel Higgs exchange}
%
\begin{equation}
	\overline{|\mathcal{M}|^2} = y_\mathrm{f}^2 y_\nu^2 \frac{(t-(m_\nu + \ms)^2)(t-m_f^2)}{(t-m_h^2)^2},
\end{equation}
In the massless limit of the external particles, this is simplified as
\dis{
	\int\frac{d \cos\theta}{2}
	\overline{|\mathcal{M}|^2} \simeq &
	N_c y_\mathrm{f}^2 y_\nu^2 \left[\frac{ \left(s+2 m_h^2\right)}{s+m_h^2}+\frac{2 m_h^2}{s}\log\frac{m_h^2 }{ m_h^2+s} \right]\\
	\simeq & N_c y_\mathrm{f}^2 y_\nu^2 \qquad  (\textrm{for}\quad s\gg m_h^2).
\label{eq:yukawa_tch}
}
%

\subsection{$\bar{f} f  \rightarrow \bar{\nu}_L \nu_s $ via $s$-channel Higgs exchange}
%
\begin{equation}
	\overline{|\mathcal{M}|^2} = N_c y_\mathrm{f}^2 y_\nu^2 \frac{(s-(m_\nu + m_{ \nu_{s_j} })^2)(s-m_f^2)}{(s-m_h^2)^2+m_h^2 \Gamma_h^2},
\end{equation}
In the massless limit of the external particles, this is simplified as
\dis{
	\int\frac{d \cos\theta}{2} \overline{|\mathcal{M}|^2} \simeq &
	N_c y_\mathrm{f}^2 y_\nu^2 \frac{ s^2 }{(s-m_h^2)^2+m_h^2 \Gamma_h^2},\\
	\simeq & N_c y_\mathrm{f}^2 y_\nu^2 \qquad  (\textrm{for}\quad s\gg m_h^2).
\label{eq:yukawa_sch}
}


\end{document}